\documentclass[a4paper,11pt]{article}
\pdfoutput=1 

\usepackage{jheppub} 

\usepackage[T1]{fontenc} 

\usepackage{mathrsfs}
\usepackage{amsmath}
\usepackage{hyperref}

\usepackage{graphicx}
\usepackage{dcolumn}
\usepackage{bm}
\usepackage{upgreek}
\def\p{\mbox{\boldmath$\displaystyle\boldsymbol{p}$}}
\def\k{\mbox{\boldmath$\displaystyle\boldsymbol{k}$}}

\def\q{\mbox{\boldmath$\displaystyle\boldsymbol{q}$}}

\def\0{\mbox{\boldmath$\displaystyle\boldsymbol{0}$}}

\def\x{\mbox{\boldmath$\displaystyle\boldsymbol{x}$}}
\def\y{\mbox{\boldmath$\displaystyle\boldsymbol{y}$}}

\renewcommand{\d}{\mathrm{d}}

\newcommand{\e}{\mathrm{e}}
\renewcommand\thefigure{\arabic{figure}}
\newcommand{\<}{\langle}
\renewcommand{\>}{\rangle}

\newcommand{\dual}[1]{\overset{\:{}^{^{{{\neg}}}}}{\smash[t]{#1}}} 

\title{\boldmath Correlators for pseudo Hermitian systems } 
\author[a]{Yao Bai}
\affiliation[a]{Department of Physics and Chongqing Key Laboratory for Strongly Coupled Physics,
Chongqing University, Chongqing 401331, China} 
\emailAdd{17381048687@163.com}
\author[b]{Ting-Long Feng}
\affiliation[b]{Xi'an Jiaotong University, Xi'an 710049, China} 
\emailAdd{arendelle.ftl@gmail.com}
\author[c]{Suro Kim}
\affiliation[c]{Korea Institute for Advanced Study, Seoul 02455, Republic of Korea} 
\emailAdd{surokim@kias.re.kr}
\author[d]{Cheng-Yang Lee}
\affiliation[d]{Center for theoretical physics, College of Physics, Sichuan University, \\Chengdu, 610064, China } 
\emailAdd{cylee@scu.edu.cn} 
\author[e]{Lei-Hua Liu}
\affiliation[e]{Department of Physics, College of Physics, Mechanical and Electrical Engineering, Jishou University, Jishou 416000, China} 
\emailAdd{liuleihua8899@hotmail.com}  
\author[a]{Wangping Zhao}
\emailAdd{20234489@stu.cqu.edu.cn} 
\author[a]{Siyi Zhou}
\emailAdd{siyi@cqu.edu.cn} 

\abstract{Pseudo-Hermitian system is a class of non-Hermitian system with Hamiltonian satisfying the condition $\eta^{-1}H^\dagger\eta=H$. We develop the in-in and Schwinger Keldysh formalism to calculate cosmological correlators for pseudo-Hermitian systems. We study a model consists of massive symplectic fermions coupled to the primordial curvature perturbation. The three-point function for the primordial curvature perturbation is computed up to one-loop and compared to earlier work where the loop correction comes from a massive scalar boson. The two results differ by a minus sign. Therefore, the one loop correction to the three-point function cannot be used to distinguished scalar bosons and symplectic fermions. To conclude, we discuss possibilities where the scalar bosons and symplectic fermions may be distinguished.}

\begin{document} 
\maketitle
\flushbottom

\section{Introduction}

Unitarity has played a crucial rule in fundamental physics because they ensure conservation of probability and information for Hermitian theories. As for non-Hermitian theories, they have found applications in describing open and dissipative systems~\cite{Ashida:2020dkc}. In such cases, dissipations and noises are described by non-Hermitian terms in the Hamiltonians~\cite{Liu:2018kfw,Brauner:2022rvf,Hongo:2018ant,Salcedo:2024smn}.

Amongst non-Hermitian theories, those with \textit{pseudo Hermitian} Hamiltonians are particularly important. They are equipped with an inner-product that preserves time translation symmetry. Therefore, when the spectrum of pseudo Hermitian Hamiltonians are real, they may describe closed systems~\cite{Mostafazadeh:2001jk,Mostafazadeh:2008pw}. How this is achieved will be discussed in sec.~\ref{pseudoH}. 

To date, various pseudo Hermitian quantum field theories have been studied in the literature with different motivations ranging from condensed matter physics~\cite{Kapit:2008dp,LeClair_2007}, dS/CFT correspondence~\cite{Fei:2015kta,Sato:2015tta,Anninos:2011ui,Ng:2012xp,Chang:2013afa,Ryu:2022ffg} and particle physics~\cite{Alexandre:2019jdb,Alexandre:2020wki,Alexandre:2020gah,Alexandre:2022uns,Mason:2023sgr,Alexandre:2023afi,Ahluwalia:2022zrm,Lee:2023aip,Sablevice:2023odu,Ahluwalia:2023slc}. Building upon these works, we develop the in-in and Schwinger Keldysh (SK)  formalism~\cite{Schwinger:1960qe,PhysRevD.33.444,Weinberg:2005vy,Chen:2016nrs,Chen:2017ryl,Donath:2024utn} for pseudo Hermitian systems in curved space-time. Both formalisms are important tools used by physicists to compute observables in cosmology~\cite{Weinberg:2005vy,Wang:2013zva,Chen:2017ryl}. Different from the in-out formalism which is used to compute the $S$-matrix in Minkowski space-time, cosmological observables are usually higher point equal time correlation functions. These include the power spectrum, bispectrum or trispectrum which are evaluated at the end of inflation.

For pseudo Hermitian Hamiltonians, it is necessary to introduce the $\eta$ product between states in the Hilbert space to preserve time translation symmetry~\cite{Mostafazadeh:2001jk,Mostafazadeh:2008pw}. Utilizing the $\eta$ product, we derive the expectation value of operators in both the in-in and SK formalism. Since the pseudo Hermitian theories of interest to us are local and respect Lorentz symmetry in Minkowski space-time, the expectation value of pseudo Hermitian operators take the same form as their Hermitian counterparts. Therefore, the Wick theorem and results from path-integral are also valid for pseudo Hermitian theories. 

For the purpose of computing correlators, there are no essential differences between pseudo Hermitian and Hermitian field theories provided that we have the following caveat - the pseudo Hermitian Hamiltonian must have a real spectrum. What happens when the spectrum is complex is beyond the scope of the present work. Fortunately for us, the pseudo Hermitian Hamiltonian that we study here has a real spectrum. 

To illustrate the formalism, we consider a model of symplectic fermions~\cite{LeClair_2007} coupled to the primordial curvature perturbation. Using the SK formalism, we compute the three-point function for the primordial curvature perturbation up to one-loop. We compare our result to earlier work~\cite{Li:2019ves} where the loop correction comes from the massive scalar bosons and that found they differ by a minus sign (interactions in both models take the same form). This difference in the minus sign cannot distinguish the bosons from the symplectic fermions. In sec.~\ref{conclusion}, we discuss other ways in which they can be distinguished.


This paper is organized as follows. In sec.~\ref{pseudoH}, we review the basics of pseudo Hermitian theory in Minkowski space-time and showed that it can also be formulated in curved space-time. In sec.~\ref{inin}, we develop the in-in and SK formalism for pseudo Hermitian systems. In sec.~\ref{ps}, we study a model consists of a massive symplectic fermion coupled to the primordial curvature perturbation. Conclusion and outlook are given in sec.~\ref{conclusion}. 

\section{Pseudo Hermitian Hamiltonians}\label{pseudoH}

In this section, we review the formalism of pseudo Hermitian Hamiltonians in Minkowski space-time extend it to curved space-time. This sets the scene for the next section where we develop the in-in and SK formalism in curved space-time.

\subsection{Minkowski space-time}\label{mink}

Let $H$ be a Hamiltonian and $\#$ be the pseudo Hermitian conjugation operator. The action of $\#$ on $H$ is defined as
\begin{equation}
    H^{\#}\equiv \eta^{-1}H^{\dag}\eta ~.\label{eq:defn}
\end{equation}
where $\eta$ is an operator to be determined. We require $H^{\#\#}=H$ so $\eta$ must be Hermitian. If the Hamiltonian is Hermitian, then~(\ref{eq:defn}) is trivial because it is a similarity transformation taking $H$ to $\eta^{-1}H\eta$ which is also Hermitian. What we are interested is the class of non-Hermitian Hamiltonians that are \textit{pseudo Hermitian}, satisfying~\cite{Mostafazadeh:2001jk,Mostafazadeh:2008pw}
\begin{equation}
	H^{\#}=H. \label{pseudo-hermitians}
\end{equation}
The spectrum of pseudo Hermitian Hamiltonians can be real or complex with non-vanishing imaginary parts.\footnote{Unless otherwise stated, from here onwards, complex eigenvalue means eigenvalue with non-vanishing imaginary part.} For the latter possibility, they come in complex conjugate pairs with equal multiplicity~\cite{Mostafazadeh:2001jk}. We review the proof here. Let $|\alpha^{(n)}\rangle$ be an eigenstate with complex eigenvalue $E_{\alpha}$
\begin{equation}
H|\alpha^{(n)}\rangle=E_{\alpha}|\alpha^{(n)}\rangle~,\label{eq:Hb}
\end{equation}
where $n=1,\cdots,N$ is the number of degenerate states with the same eigenvalue.
Taking the Hermitian conjugate of~(\ref{eq:Hb}), we see that $E^{*}_{\alpha}$ is an eigenvalue of $H^{\dag}$ with multiplicity $N$. Multiply~(\ref{eq:Hb}) from the left by $\eta$ and using~(\ref{pseudo-hermitians}), we obtain
\begin{equation}
H^{\dag}\eta|\alpha^{(n)}\rangle=E_{\alpha}\eta|\alpha^{(n)}\rangle.
\end{equation}
Therefore, both $E_{\alpha}$ and $E^{*}_{\alpha}$ are eigenvalues of $H^{\dag}$ with equal multiplicity. Since $H$ and $H^{\dag}$ have the same spectrum, the proof is complete.

For systems described by pseudo Hermitian Hamiltonians, the Hermitian inner-product that is used in unitary quantum mechanics is inadequate because it does not preserve time translation symmetry. To be precise, let $|\alpha\rangle$ and $|\beta\rangle$ be two states in the system. The time evolution for $|\alpha\rangle$ and $\langle\beta|$ are given by\footnote{Since the Hamiltonian is non-Hermitian, one is justified to ask why the time evolution of $|\alpha\rangle$ is generated by $H$ and not $H^{\dag}$. In fact, both are physically equivalent. This is because the definition of pseudo Hermiticity can be understood as a similarity transformation between $H^{\dag}$ and $H$. That is, if we evolve $|\alpha\rangle$ using $H$, then state $\eta|\alpha\rangle$ will evolve under $H^{\dag}$.}
\begin{equation}
    |\alpha(t)\rangle=e^{-iHt}|\alpha\rangle,\quad
    \langle\beta(t)|=\langle\beta| e^{iH^{\dag}t}.\label{trans}
\end{equation}
The invariant inner-product is the $\eta$ product~\cite{Mostafazadeh:2001jk}
\begin{equation}    \langle\beta|\alpha\rangle_{\eta}\equiv\langle\beta|\eta|\alpha\rangle.\label{eta_prod}
\end{equation}
Substituting~\eqref{trans} into~\eqref{eta_prod}, we obtain
\begin{align}
\langle\beta(t)|\alpha(t)\rangle_{\eta}&=\langle\beta|e^{iH^{\dag}t}\eta e^{-iHt}|\alpha\rangle_{\eta} \nonumber\\
&=\langle\beta|\alpha\rangle_{\eta}~,\label{inv_prod}
\end{align}
so the $\eta$ product is invariant under time translation. On the second line of~(\ref{inv_prod}), we have used the identity $e^{iH^{\dag}t}\eta e^{-iHt}=\eta$. 

The $\eta$ product and the definition of pseudo Hermitian Hamiltonian allow us to derive important properties of the eigenstates. Let $|\alpha\rangle$ and $|\beta\rangle$ be eigenstates of $H$ with eigenvalues $E_{\alpha}$ and $E_{\beta}$. Using~(\ref{pseudo-hermitians}), we find
\begin{align}
    0&=\langle\beta|(H^{\dag}\eta-\eta H)|\alpha\rangle\nonumber\\
    &=(E^{*}_{\beta}-E_{\alpha})\langle\beta|\alpha\rangle_{\eta}.\label{ee}
\end{align}
In the special case where $\alpha=\beta$, we have
\begin{equation}
    (E^{*}_{\alpha}-E_{\alpha})\langle\alpha|\alpha\rangle_{\eta}=0.
\end{equation}
Therefore, an eigenstate has real eigenvalue when its $\eta$ norm is non-vanishing. If an eigenstate has complex eigenvalue, then its $\eta$ norm vanishes.

To derive the completeness relation, let us denote $|\beta^{*}\rangle$ as an eigenstate of $H$ with eigenvalue $E^{*}_{\beta}$. Equation~(\ref{ee}) then becomes
\begin{equation}
    0=(E_{\beta}-E_{\alpha})\langle\beta^{*}|\alpha\rangle_{\eta}.
\end{equation}
Therefore, when $E_{\beta}=E_{\alpha}$ the inner-product $\langle\alpha^{*}|\alpha\rangle_{\eta}$ may be non-vanishing. Without knowing $\eta$, we cannot compute $\langle\alpha^{*}|\alpha\rangle_{\eta}$. Nevertheless knowing that $\langle\beta^{*}|\alpha\rangle_{\eta}$ vanishes when $E_{\beta}\neq E_{\alpha}$ is allow for us to infer
\begin{equation}
    \langle\beta^{*}|\alpha\rangle_{\eta}=\delta(\beta-\alpha).\label{eq:ba}
\end{equation}
Here we take the $\eta$ product to be positive-definite. This may not be true for all pseudo Hermitian theories, but for the particular field theories that we are working with here, we can always define the $\eta$ product to be positive-definite. Therefore, the completeness relation takes the form
\begin{equation}
    \sum_{\alpha}|\alpha\rangle\langle\alpha^{*}|\eta=I.
\end{equation}

To develop the in-in and SK formalism in curved space-time, it is instructive to review the definitions of the \textit{in} and \textit{out} states in Minkowski space-time describing scattering processes. We take the full Hamiltonian to be
\begin{equation}
    H=H_{0}+V, \label{eq:H0}
\end{equation}
where $H_{0}$ and $V$ are the free and interacting parts. Splitting the Hamiltonian via~(\ref{eq:H0}) simplifies the analysis. This is because for any physically well-defined theories in Minkowski space-time, the free Hamiltonians must be Hermitian and have positive-definite real spectrum. Furthermore, it can be shown that $H$ and $H_{0}$ have the same spectrum so that $|\alpha^{*}\rangle=|\alpha\rangle$~\cite{Lee:2024}.

Let $|\alpha_{0}\rangle$ be the free state that evolves under $H_{0}$ and $|\alpha_{-}\rangle$ and $|\alpha_{+}\rangle$ be the in and out states that evolve under $H$. We take the scattering to occur around the time $t=0$ so in the limit $t\rightarrow\pm\infty$, the states are free. Therefore, $|\alpha_{\pm}\rangle$ and $|\alpha_{0}\rangle$ related by
\begin{equation}
    \lim_{t\rightarrow\pm\infty}e^{-iHt}|\alpha_{\pm}\rangle=\lim_{t\rightarrow\pm\infty}e^{-iH_{0}t}|\alpha_{0}\rangle,\label{eq:pm1}
\end{equation}
which maybe rewritten as
\begin{equation}
    |\alpha_{\pm}\rangle=\Omega_{\pm}|\alpha_{0}\rangle,\label{eq:pm2}
\end{equation}
where 
\begin{equation}
    \Omega(t)=e^{iHt}e^{-iH_{0}t},\quad \Omega_{\pm}=\lim_{t\rightarrow\pm\infty}\Omega(t).
\end{equation}
Since $H$ is pseudo Hermitian, $\Omega^{-1}$ is given by its pseudo Hermitian conjugate 
\begin{equation}
\Omega^{-1}(t)=\eta^{-1}\Omega^{\dag}(t)\eta.\label{eq:Omega}
\end{equation}
Using~(\ref{eq:Omega}), we find that the $\eta$ product for the in and out states are identical to the $\eta$ product of the free state
\begin{align}
    \langle\beta_{\pm}|\alpha_{\pm}\rangle&=\langle\beta_{0}|\alpha_{0}\rangle_{\eta},\label{eq:prod_pm}
\end{align}
so its completeness relation is given by
\begin{equation}
    \sum_{\alpha}|\alpha_{\pm}\rangle\langle\alpha_{\pm}|\eta=I.\label{eq:cr}
\end{equation}

\subsection{Curved space-time}

Pseudo Hermitian systems in curved space-time is similar to its formulation in Minkowski space-time. Therefore, the theorems presented in sec.~\ref{mink} concerning the properties of the eigenvalues and eigenstates are also valid in curved space-time.

The in, out and the free states can also be defined in curved space-time. But because Hamiltonian is now time-dependent, the in and out state are not related to the free states via~(\ref{eq:pm1}-\ref{eq:Omega}). Here we will only deal with the in states so we denote them as $|\alpha\rangle$. Given two in states at time $t_{0}$ and $t$ where $t_{0}\leq t$, we have
\begin{equation}
|\alpha_,t\rangle=U(t,t_{0})|\alpha,t_{0}\rangle,
\end{equation}
where $U$ is the evolution operator satisfying the Schr\"{o}dinger equation
\begin{equation}
    i\frac{d}{dt}U(t,t_{0})=H(t)U(t,t_{0}), \label{eq:sch}
\end{equation}
with the initial condition 
\begin{equation}
    U(t_{0},t_{0})=1.\label{eq:ini}
\end{equation}
In~(\ref{eq:sch}), $H$ is the full Hamiltonian. In cosmology, the limit $t_{0}\rightarrow-\infty$ means that the wavelengths are inside the horizon and the in state becomes the free state so we have
\begin{equation}
    |\alpha,t\rangle=\lim_{t_{0}\rightarrow-\infty}U(t,t_{0})|\alpha_{0},t_{0}\rangle. \label{eq:alpha_in}
\end{equation}
In a pseudo Hermitian system, $H$ is pseudo Hermitian so $U$ satisfies the generalized unitarity condition
\begin{equation}
    U^{-1}(t,t_{0})=\eta^{-1}U^{\dag}(t,t_{0})\eta.\label{eq:pH_U}
\end{equation}
Therefore, the $\eta$ product for the in state is equal to the $\eta$ product of the free state. Their inner-products and completeness relations are also given by~(\ref{eq:prod_pm}-\ref{eq:cr}). The solution for $U$ is presented in the next section.





\section{Correlators}\label{inin} 

Multi-point correlation function of operators are important in early universe cosmology, especially for studying the quantum fluctuations generated during inflation. For an operator $Q$, the quantity we wish to compute is the expectation value
\begin{equation}
    \langle Q(\tau)\rangle_{\eta}\equiv\langle\Omega|\eta Q(\tau)|\Omega\rangle, \label{eq:vev_Q}
\end{equation}
where $|\Omega\rangle$ is the in-vacuum. This definition ensures that $\langle Q\rangle_{\eta}$ is invariant time translation generated by pseudo Hermitian Hamiltonians. When $Q$ is pseudo Hermitian, then~(\ref{eq:vev_Q}) is real because $\eta Q$ is Hermitian. If $Q$ is Hermitian, we require it to commute with $\eta$ so that $\eta Q$ remains Hermitian.

The in-in and SK formalisms~\cite{Schwinger:1960qe,PhysRevD.33.444,Weinberg:2005vy,Chen:2016nrs,Chen:2017ryl,Donath:2024utn} have been developed to compute correlators for unitary quantum field theories. Here we show that both formalisms can be applied to pseudo Hermtian field theories.

\subsection{In-in formalism}

The derivations to the relevant formulae for the pseudo Hermitian in-in formalism are by in large, the same as the Hermitian field theories with the exception that pseudo Hermitian Hamiltonians may have complex eigenvalues. For most part, the results are identical to those given in~\cite{Weinberg:2005vy,Wang:2013zva} so we will simply present them without proof. 

Let $\Phi_{a},\Pi_{a}$ be the canonical operators in the Heisenberg picture and $\phi_{a},\pi_{a}$ be the perturbations from the classical background in interacting picture. Their evolutions from time $t_{0}$ to $t$ are given by
\begin{align}
    \Phi_{a}(t,\x)&=U(t,t_{0})\Phi_{a}(t_{0},x)U^{-1}(t,t_{0}),\\
    \Pi_{a}(t,\x)&=U(t,t_{0})\Pi_{a}(t_{0},x)U^{-1}(t,t_{0}),
\end{align}
and
\begin{align}
    \phi_{a}(t,\x)&=U_{0}(t,t_{0})\phi_{a}(t_{0},x)U^{-1}_{0}(t,t_{0}),\\
    \pi_{a}(t,\x)&=U_{0}(t,t_{0})\pi_{a}(t_{0},x)U^{-1}_{0}(t,t_{0}),
\end{align}
where $U_{0}$ is the evolution operator for the free fields. In the limit $t_{0}\rightarrow-\infty$, they satisfy the initial conditions
\begin{equation}
    \Phi_{a}(t_{0},\x)=\phi_{a}(t_{0},\x),\quad
    \Pi_{a}(t_{0},\x)=\pi_{a}(t_{0},\x).
\end{equation}
and 
\begin{equation}
    U_{0}(t_{0},t_{0})=U(t_{0},t_{0})=1.
\end{equation}
Taking the Hamiltonian to be
\begin{equation}
    H=H_{0}+V,
\end{equation}
where $H_{0}$ and $H_{I}$ are the free and interacting parts, the solutions for $U$ and $U_{0}$ are
\begin{align}
       U(t,t_{0})&=U_{0}(t,t_{0})F(t,t_{0}),
\end{align}
where
\begin{align}
 U_{0}(t,t_{0})&=T\exp\left[-i\int^{t}_{t_{0}}dt'H_{0}(t')\right]\label{eq:U0soln}, \\
    F(t,t_{0})&=T\exp\left[-i\int^{t}_{t_{0}}dt'V(t')\right],\label{eq:F}
\end{align}
with $T$ being the time-ordering operator such that in the power series expansion the time argument of the fields increase from right to left. Since $H_{I}$ is pseudo Hermitian, $F^{-1}$ is 
\begin{equation}
    F^{-1}(t,t_{0})=\eta^{-1}F^{\dag}(t,t_{0})\eta.\label{eq:F_g}
\end{equation}
We then have
\begin{align}
    \Phi_{a}(t)&=F^{-1}(t,t_{0})\phi_{a}(t)F(t,t_{0}), \\
    \Pi_{a}(t)&=F^{-1}(t,t_{0})\pi_{a}(t)F(t,t_{0}).
\end{align}
Therefore, given a $Q$ that is a function of $\Phi_{a}$ and $\Pi_{a}$, it is related to $Q^{I}$ in the interacting picture via
\begin{equation}
    Q(t)=F^{-1}(t,t_{0})Q^{I}(t)F(t,t_{0}).
\end{equation}

To compute~(\ref{eq:vev_Q}), we need the relation between the in vacuum $|\Omega\rangle$ and the free vacuum $|0\rangle$. In curved space-time, the free Hamiltonian $H_{0}$ may not be Hermitian so we have to use the completeness relation~(\ref{eq:cr}). We find
\begin{align}
	e^{-iH(t-t_0)}|0\>=&\sum_{\alpha}\left[e^{-i H(t-t_0)}|\alpha\>\<\alpha^{*}|0\>_{\eta}\right]\nonumber\\
 &=|\Omega\>\<\Omega|0\>_{\eta}+\sum_{\alpha\neq\Omega}\left[e^{-iE_{\alpha}(t-t_0)}|\alpha\>\<\alpha^{*}|0\>_{\eta}\right].\label{eq:expH0}
\end{align}
If the eigenvalues of $H$ are real and positive, then by analytically extend $(t-t_{0})$ to the complex plane
\begin{equation}
    (t-t_{0})\rightarrow(\widetilde{t}-\widetilde{t}_{0})(1-i\epsilon),
\end{equation}
the contribution from the in vacuum  $|\Omega\rangle$ dominates in the limit $\widetilde{t}_{0}\rightarrow-\infty$. If $H$ has complex eigenvalues, then we need to examine the inner-product $\langle\alpha^{*}|0\rangle_{\eta}$. Let us examine this term in Minkowski space-time and then in curved space-time. 

As we have discussed in sec.~\ref{mink}, for any physically well-defined theories in Minkowski space-time, the free states and in states have the same spectrum with respect to the free Hamiltonian $H_{0}$ and the full Hamiltonian $H$ respectively. Now, since $H_{0}$ is always Hermitian, its eigenvalues are real so it follows that the spectrum of $H$ is real. Therefore, in Minkowski space-time, there are no in states with complex eigenvalues so in~(\ref{eq:expH0}), the dominant state is $|\Omega\rangle$ as $\widetilde{t}_{0}\rightarrow-\infty$.

In curved space-time, both the free and full Hamiltonians can be pseudo Hermitian so in general, their spectra may be complex. If the spectra are complex, then it is necessary to compute $\langle\alpha^{*}|0\rangle_{\eta}$. Using the fact that the $\eta$-product is invariant under time translation, we have $\langle\alpha^{*}|0\rangle_{\eta}=\langle\alpha^{*},t|0,t\rangle_{\eta}$ for all $t$. Therefore, we can take the limit $t\rightarrow-\infty$ to obtain
\begin{align}
    \langle\alpha^{*}|0\rangle_{\eta}&=\lim_{t\rightarrow-\infty}e^{iE_{\alpha}t}\langle\alpha^{*}_{0}|\eta e^{-iHt}|0\rangle.\label{eq:a_star_0}
\end{align}
Because the $\eta$-product is translation invariant, $\langle\alpha^{*}|0\rangle_{\eta}$ cannot be divergent as $t\rightarrow-\infty$. If $e^{-iHt}|0\rangle$ does not contain states with complex eigenvalues then~$\langle\alpha^{*}|0\rangle_{\eta}$ vanishes so $|\Omega\rangle$ dominates in~(\ref{eq:expH0}) as $\widetilde{t}_{0}\rightarrow-\infty$. On the other hand, if $e^{-iHt}|0\rangle$ yields a state proportional to $|\alpha_{0}\rangle$ where $E_{\alpha}$ is complex, then we must have $\langle\alpha^{*}|0\rangle_{\eta}=\langle\alpha^{*}_{0}|\alpha_{0}\rangle_{\eta}$. In this case, the above prescription to relate $|\Omega\rangle$ and $|0\rangle$ becomes inadequate. But fortunately for us, as we will show in sec.~\ref{ps}, the free and full Hamiltonians of symplectic fermions have no complex eigenstates so $\langle\alpha^{*}_{0}|0\rangle_{\eta}$ vanishes. Therefore, in the limit $\widetilde{t}_{0}\rightarrow-\infty$, we can neglect the last term in~(\ref{eq:expH0}) to obtain
\begin{equation}
 \e^{-i H(t-t_0)}|\Omega\>=\frac{\e^{-i H(t-t_0)}|0\>}{\<\Omega|0\>},
\end{equation}
and hence
\begin{equation}\label{A.5}
	F(t,t_0)|\Omega\>=\frac{F(t,t_0)|0\>}{\<\Omega|0\>},\quad
    \<\Omega|F^{\dag}(t,t_{0})=\frac{\<0|F^{\dag}(t,t_{0})}{\<0|\Omega\>}.
\end{equation}
Therefore, we obtain
\begin{align}
    \langle Q(t)\rangle_{\eta}=&\langle\Omega|\eta F^{-1}(t,t_{0})Q^{I}(t)F(t,t_{0})|\Omega\rangle\nonumber\\
    =&\langle0|\eta F^{-1}(t,t_{0})Q^{I}(t)F(t,t_{0})|0\rangle,
\end{align}
where on the second line, we have used~(\ref{eq:F_g}) and~(\ref{A.5}).

\subsection{Schwinger Keldysh formalism}

To compute the cosmological correlators, it is more convenient to use the SK formalism which utilizes the path-integral.  Here we mostly follow the presentation given in~\cite{Chen:2017ryl}. 

An important difference between Hermitian and pseudo Hermitian field theories is the completeness relation. Following the discussions in sec.~\ref{pseudoH}, the completeness relation for pseudo Hermitian theories take the form
\begin{equation}
    I=\sum_{\alpha}|O_{\alpha}\rangle\langle O_{\alpha}|\eta,
\end{equation}
so we have
\begin{equation}
    \langle Q(\tau)\rangle_{\eta}=\sum_{\alpha}\langle\Omega|\eta|O_{\alpha}\rangle\langle O_{\alpha}|\eta Q(\tau)|\Omega\rangle.\label{eq:q_eta}
\end{equation}

For the canonical fields and conjugate momenta eigenstates in the pseudo Hermitian system, we have
\begin{equation}
    I=\int d\phi|\phi(\tau,\x)\rangle\langle\phi(\tau,\x)|\eta=\int d\pi|\pi(\tau,\x)\rangle\langle\pi(\tau,\x)|\eta.
\end{equation}
Their inner-products are the $\eta$-products are
\begin{align}
    \langle\phi'(\tau,\x)|\phi(\tau,\x)\rangle_{\eta}&=\delta(\phi'(\tau,\x)-\phi(\tau,\x)),\label{eq:pp} \\
    \langle\pi'(\tau,\x)|\pi(\tau,\x)\rangle_{\eta}&=\delta(\pi'(\tau,\x)-\pi(\tau,\x)),
\end{align}
and
\begin{equation}
    \langle\phi(\tau,\x)|\pi(\tau,\x)\rangle_{\eta}=\exp\left[i\int d^{3}x\phi(\tau,\x)\pi(\tau,\x)\right].\label{eq:phi_pi_ip}
\end{equation}
The right-hand side of~(\ref{eq:pp}-\ref{eq:phi_pi_ip}) are identical to inner-products for Hermitian theories. These expressions will be justified in the next section when we consider the symplectic fermions. So apart from replacing Hermitian inner-product with the $\eta$ product, the path-integral for pseudo Hermitian system is identical to the Hermitian system. Therefore, we may follow~\cite{Chen:2017ryl} and define the generating functional $Z[J_{+},J_{-}]$
\begin{align}
    &Z[J_{+},J_{-}]\nonumber\\
    &=\int D\phi_{+}D\phi_{-} \exp\left\{i\int^{\tau_{f}}_{\tau_{0}}d\tau\int d^{3}x\Big[\mathcal{L}[\phi_{+}]-\mathcal{L}\left[\phi_{-}\right]+\left(J_{+}\phi_{+}-J_{-}\phi_{-}+\text{$\#$ conjugate terms}\right)\Big]\right\}.\label{eq:Zpm}
\end{align}
The expectation value for $\phi(\tau,\x_{1})\cdots\phi(\tau,\x_{N})$ is
\begin{align}
    \langle\phi(\tau,\x_{1})\cdots\phi(\tau,\x_{N})\rangle_{\eta}=&\int D\phi_{+}D\phi_{-}
    \left[\phi_{+}(\tau,\x_{1})\cdots\phi_{+}(\tau,\x_{N})\right]\nonumber\\
    &\times\exp\left[i\int^{\tau_{f}}_{\tau_{0}}d\tau\int d^{3}x\left(\mathcal{L}[\phi_{+}]-\mathcal{L}\left[\phi_{-}\right]\right)\right].
\end{align}

\section{Cosmological signatures for symplectic fermions}\label{ps}

Symplectic fermion is a theory of anti-commuting complex scalar fields proposed by LeClair and Neubert~\cite{LeClair_2007}. In Minkowski space-time, the theory has the following feature. It evades the spin-statistics theorem because its field adjoint is pseudo Hermitian while respecting locality and Lorentz symmetry. 


Here we study the cosmological collider signals produced by the symplectic fermions interacting with the primordial curvature perturbation $\zeta$ in the inflationary background
\begin{equation}
    ds^{2}=-dt^{2}+e^{2(Ht+\zeta)}d\x\cdot d\x,
\end{equation}
where $H$ is the Hubble parameter. We present the theory with the above background, establish its pseudo Hermiticity and then compute the three-point function for $\zeta$ using the SK formalism.

\subsection{Symplectic fermions}

The action for symplectic fermions in curved space-time is~\cite{LeClair_2007,Lee:2023aip}
\begin{align} 
S&=-\int \d^{4}x \sqrt{-g}\left[g^{\mu\nu}(\partial_\mu\dual{\sigma})(\partial_\nu\sigma)+m^{2}\dual{\sigma}\sigma\right].
\end{align} 
In the momentum space,
\begin{align}
    \sigma(x)&=\int\frac{d^{3}k}{(2\pi)^{3}}e^{i\boldsymbol{k\cdot x}}\sigma(\k,t),\label{eq:phix_i}\\
    \dual{\sigma}(x)&=\int\frac{d^{3}k}{(2\pi)^{3}}e^{-i\boldsymbol{k\cdot x}}\dual{\sigma}(\k,t),\label{eq:dphix_i}
\end{align}
where 
\begin{align}  
\sigma(\k,t)&=v(k,t)a(\k)+v^*(k,t)b^\dag(-\k),\label{eq:sigma} \\ 
\dual{\sigma}(\k,t)&=v^{*}(k,t)a^\dag_{i}(\k)+v(k,t) b(-\k),\label{eq:dsigma}
\end{align} 
with $v$ being the mode function and $a$ and $b^{\dag}$ are the annihilation and creation operators. At the zeroth order of the slow-roll parameter $O(\epsilon^{0})$, the mode function $v_k$ satisfies the following equation of motion. 
\begin{align}
\ddot v_k+ 3 H \dot v_k +\frac{k^2}{a^2} v_k +m^2 v_k^2 = 0. \label{eq:v_k}
\end{align}
The solution to~(\ref{eq:v_k}), which in the limit $\tau\rightarrow-\infty$ becomes the oscillatory phase $e^{-ik\tau}$ is
\begin{align}
v_{k}(\tau)=e^{-\pi\mu/2}\frac{\sqrt{\pi}H}{2}(-\tau)^{3/2}H^{(1)}_{i\mu}(-k\tau),
\end{align}
where $\mu\equiv \sqrt{m^2/H^2-9/4}$ and $H^{(1)}_{i\mu}$ is the Hankel function of the first kind. Here we focus on the case where $m>\frac{3}{2}H$.

The fields are \textit{fermionic} and \textit{pseudo Hermitian} when the annihilation and creation operators satisfy~\cite{Ryu:2022ffg}
\begin{align}
    \left\{ a(\bm{k}), a^\dagger(\bm{k}')\right\}&=+(2\pi)^3 \delta^3 (\bm{k}-\bm{k}'), \\
    \left\{ b(\bm{k}), b^\dagger(\bm{k}')\right\}&=-(2\pi)^3 \delta^3 (\bm{k}-\bm{k}'). \label{eq:bb}
\end{align}
The minus sign in~(\ref{eq:bb}) yields states with negative norm but it can be removed by introducing the $\eta$ product. By demanding the action of $\eta$ leaves the free vacuum state $|0\rangle$ invariant and that it commute and anti-commute with $a$ and $b$, we find
\begin{equation}
    \eta=\exp\left[-i\pi\int d^{3}k\,b^{\dag}(\k)b(\k)\right].\label{eq:eta_i}
\end{equation}
The $\eta$ product for single state created by $a^{\dag}$ and $b^{\dag}$ are positive-definite
\begin{equation}
    \langle\k,a|\k',a\rangle_{\eta}=\langle\k,b|\k',b\rangle_{\eta}=(2\pi)^{3}\delta^{3}(\k'-\k).
\end{equation}
The fields satisfy the canonical anti-commutation relations
\begin{align}
    \left\{\sigma(\tau,\x),\dual{\sigma}(\tau,\y)\right\}&=0,\\
    \left\{\sigma(\tau,\x),\pi(\tau,\y)\right\}&=i\delta^{3}(\x-\y),
\end{align}
where $\pi=\partial_{\tau}\dual{\sigma}$. The theory is pseudo Hermitian because 
\begin{equation}
    \eta^{-1}\left[\dual{\sigma}(x)\sigma(x)\right]^{\dag}\eta=\dual{\sigma}(x)\sigma(x).\label{eq:ph_phi}
\end{equation}
From~(\ref{eq:eta_i}) and using~(\ref{eq:prod_pm}), the $\eta$ product for free and the in states are given by
\begin{align}
    \langle\beta|\alpha\rangle_{\eta}&=\langle\beta_{0}|\alpha_{0}\rangle_{\eta}
    =\delta(\beta-\alpha).
\end{align}
The $\eta$-norm is non-vanishing, so the free and in states have real eigenvalues. Since all states are obtained by acting the creation operators on the vacuum, their $\eta$ norms are non-vanishing. Therefore, the eigenvalues of the full Hamiltonian are real. As the $\eta$ product is positive-definite and the fields are local, we expect the functional relations~(\ref{eq:pp}-\ref{eq:phi_pi_ip}) to hold and that the path-integral for symplectic fermions to be well-defined.

\subsection{Coupling to primordial curvature perturbation}

\begin{figure}
\centering
\includegraphics[height=2.5in]{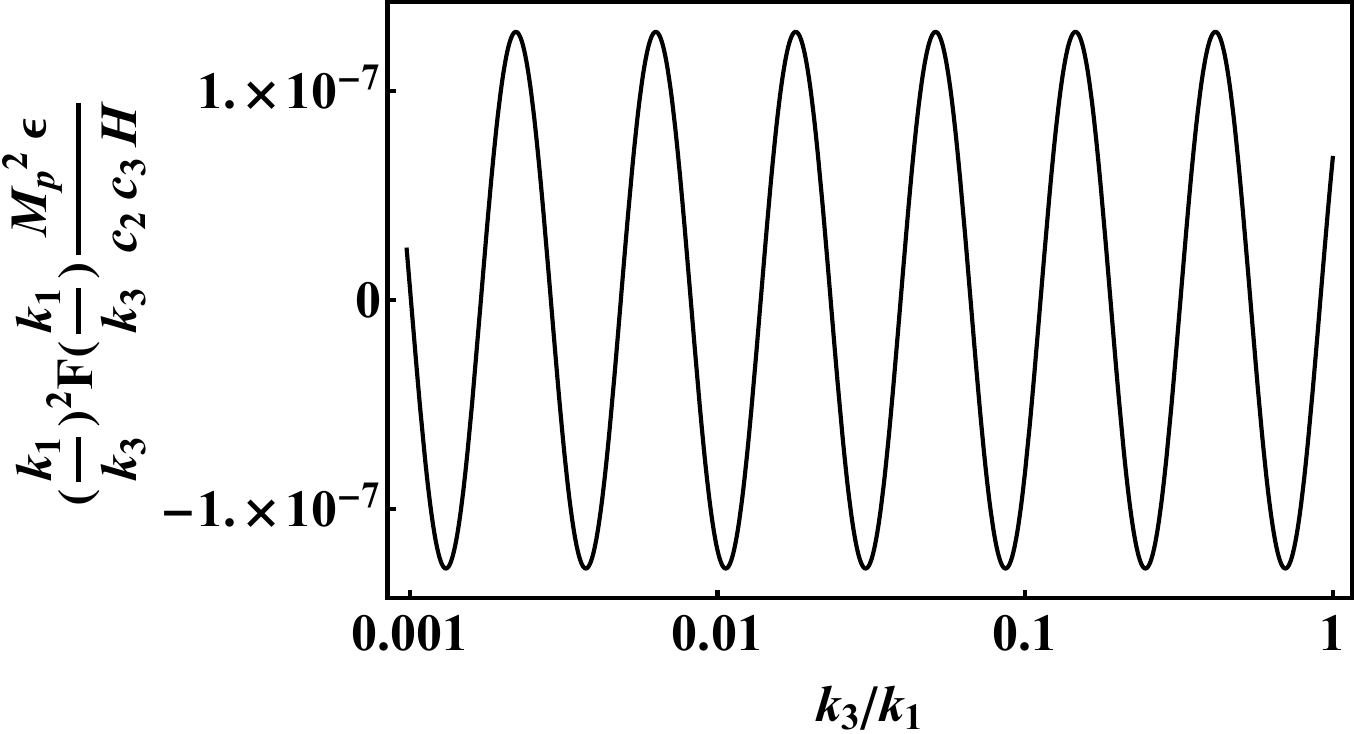}
\caption{An illustration of the non-Gaussianities in the squeezed limit for symplectic fermions.}\label{fig:genericS}
\end{figure}

We now consider the cosmological collider signals produced by the symplectic fermions interacting with the primordial curvature perturbation $\zeta$. The second order action for $\zeta$ is~\cite{Chen:2010xka,Wang:2013zva}
\begin{align}
S_\zeta=M_p^2 \int d t \frac{d^3 k}{(2 \pi)^3} \epsilon\left(a^3 \dot{\zeta}^2-k^2 a \zeta^2\right).
\end{align}
where $\epsilon\equiv-\dot{H}/H^{2}$ is the slow-roll parameter. Quantizing $\zeta$ gives
\begin{align}
\zeta_{\boldsymbol{k}}(\tau)=u_{k}(\tau) c(\k)+u^{*}_{k}(\tau)c^{\dagger}(-\k), \label{eq:zeta_k}
\end{align}
where $c$ and $c^\dagger$ are the annihilation and creation operators satisfying
\begin{align}
\left[c(\k), c^{\dagger}(\k')\right]=(2 \pi)^3 \delta^{(3)}(\k-\k').
\end{align}
The equation of motion for $\zeta$ is
\begin{align}
\ddot{\zeta}+3 H \dot{\zeta}+\frac{k^2}{a^2} \zeta=0. \label{eq:zeta}
\end{align}
Solving~(\ref{eq:zeta}) yields
\begin{align}
u_k(\tau) = \frac{H}{2\sqrt{\epsilon}M_{\rm pl}}\frac{1}{k^{3/2}}.(1+ik\tau)e^{-i k \tau}.
\end{align}
\begin{figure}
    \centering
    \includegraphics[width=0.5\linewidth]{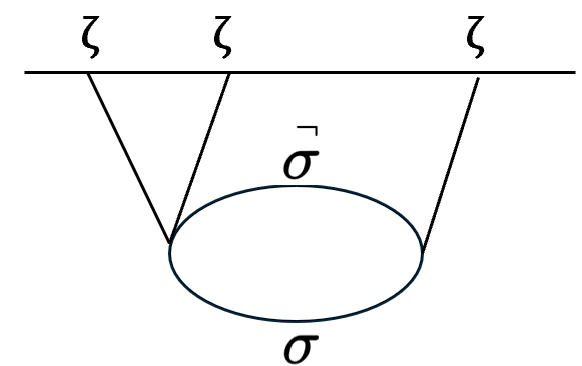}
    \caption{The leading contribution to $\langle\zeta_{\boldsymbol{k}_{1}}\zeta_{\boldsymbol{k}_{2}}\zeta_{\boldsymbol{k}_{3}}\rangle_{+}$ coming from the fermionic loop.}
    \label{fig:enter-label}
\end{figure}
We take the interactions between symplectic fermion and $\zeta$ to be
\begin{equation}
    V=V_{3}+V_{4},
\end{equation}
where
\begin{align}
V_{3}(\tau) &= c_3\int d^3 x \left[a^{3}(\tau)\zeta'(\tau,\x)\dual{\sigma}(\tau,\x)\sigma(\tau,\x)\right], \\
V_{4}(\tau) &= c_4\int d^3 x \left[a^{2}(\tau)\zeta'(\tau,\x)\zeta'(\tau,\x)\dual{\sigma}(\tau,\x)\sigma(\tau,\x)\right],
\end{align}
with $c_{3,4}$ being constants. For such an interaction, the leading contribution to the three-point function $\langle\zeta_{\boldsymbol{k}_{1}}\zeta_{\boldsymbol{k}_{2}}\zeta_{\boldsymbol{k}_{3}}\rangle_{\eta}$ is of order $O(c_{3}c_{4})$ (see fig.~\ref{z3pic}). Here $\zeta$ is Hermitian and commutes with $\eta$ so its expectation value is well-defined. For convenience, we shall ignore the subscript $\eta$ from now onwards. The three-point function has been computed in~\cite{Li:2019ves} where the massive complex scalar fields which interact with the primordial perturbation are bosonic. Here we take the massive complex scalar fields to be fermionic and compare the resulting three-point function with its bosonic counterpart. For comparison, we denote the bosonic and fermionic three-point function as
$\langle\zeta_{\boldsymbol{k}_{1}}\zeta_{\boldsymbol{k}_{2}}\zeta_{\boldsymbol{k}_{3}}\rangle_{-}$ and $\langle\zeta_{\boldsymbol{k}_{1}}\zeta_{\boldsymbol{k}_{2}}\zeta_{\boldsymbol{k}_{3}}\rangle_{+}$ respectively.  In the squeezed limit (see app.~\ref{sk_prop})
\begin{align}
\left\langle\zeta_{\mathbf{k}_1} \zeta_{\mathbf{k}_2} \zeta_{\mathbf{k}_3}\right\rangle'_{\pm}
=-\frac{c_{3}c_{4}}{4\pi^{2}H^{2}}\text{Re}\Big\{&\int^{0}_{-\infty}d\tau_{1}\int^{0}_{-\infty}d\tau_{2}\int\frac{d^{3}p}{(2\pi)^3}\int\frac{d^3q}{(2\pi)^3}\nonumber\\
\times&\delta^{3}(\p+\q-\k_{3})u(k_{1},0)u(k_{2},0)u(k_{3},0)\nonumber\\
\times&\left[\partial_{\tau_1}u^{*}(k_{3},\tau_{1})-\partial_{\tau_{1}}u(k_{3},\tau_{1})\right]
(\mp 1)\left[\Gamma^{2}(-i\mu)\left(\frac{pq\tau_{1}\tau_{2}}{4}\right)^{2i\mu}\tau_{2}+\text{c.c.}\right]\nonumber\\
\times&\left[\partial_{\tau_2}u^{*}\left(k_1,\tau_2\right)\right]\left[\partial_{\tau_{2}}u^{*}\left(k_{2},\tau_{2}\right)\right]
+(\text{even permutations in $k_{1},k_{2},k_{3}$})\Big\}.\label{eq:z_int}
\end{align}
We evaluate the integrals over $\p$ and $\q$ in~(\ref{eq:z_int}) by rewriting it as
\begin{equation}
    \int\frac{d^{3}p}{(2\pi)^3}\int\frac{d^3q}{(2\pi)^3}\delta^{3}(\p+\q-\k_{3})(pq)^{2i\mu}=
    \int\frac{d^{3}x}{(2\pi)^{3}} \int\frac{d^{3}p}{(2\pi)^3}\int\frac{d^3q}{(2\pi)^3}e^{i\boldsymbol{(p+q-k_{3})\cdot x}}(pq)^{2i\mu}.
\end{equation}
Using the Fourier transform~\cite[pg.~363]{Gelfand:1964}
\begin{equation}
    \int\frac{d^{n}x}{(2\pi)^{n}}\,e^{i\boldsymbol{x\cdot k}}k^{\lambda}=2^{\lambda}\pi^{-n/2}\frac{\Gamma\left(\frac{\lambda+n}{2}\right)}{\Gamma\left(-\frac{\lambda}{2}\right)}k^{-n-\lambda},
\end{equation}
we obtain
\begin{equation}
    \int\frac{d^{3}p}{(2\pi)^3}\int\frac{d^3q}{(2\pi)^3}\delta^{3}(\p+\q-\k_{3})(pq)^{2i\mu}=2^{-3/2}(2\pi)^{-9/2}\left[\frac{\Gamma(\frac{3}{2}+i\mu)}{\Gamma(-i\mu)}\right]^{2}\frac{\Gamma\left(-\frac{3}{2}-2i\mu\right)}{\Gamma(3+2i\mu)}k^{4i\mu+3}.
\end{equation}
Performing the time integral yields
\begin{align}
 \left\langle\zeta_{\mathbf{k}_1} \zeta_{\mathbf{k}_2} \zeta_{\mathbf{k}_3}\right\rangle'_{\pm}
=&\mp\text{Re}\Big\{\left[\frac{8c_{2}c_{3}H^{5}}{(2\pi)^{13/2}\epsilon^{3}M^{6}_{P}k_{1}k_{2}(k_{1}+k_{2})^{4}}\right]\left[\frac{k_{3}}{4(k_{1}+k_{2})}\right]^{2i\mu}\nonumber\\
&+(\text{even permutations in $k_{1},k_{2},k_{3}$})\Big\}.
\end{align}
where
\begin{equation}
    g(\mu)=\Gamma\left(\frac{3}{2}+i\mu\right)\Gamma\left(\frac{5}{2}+i\mu\right)\Gamma\left(-\frac{3}{2}+2i\mu\right)\Gamma(2-2i\mu)\sinh^{2}(\pi\mu).
\end{equation}
Therefore, the bosonic and fermionic bispectrum differs by a sign. We rewrite it as
\begin{align}
\left\langle\zeta_{\mathbf{k}_1} \zeta_{\mathbf{k}_2} \zeta_{\mathbf{k}_3}\right\rangle' \equiv (2\pi)^4 \frac{P_\zeta^2}{(k_1k_2k_3)^2} F_{\pm} \bigg(\frac{k_1}{k_3},\frac{k_2}{k_3}\bigg),
\end{align}
where $F$ is the shape function
\begin{align}
F_{\pm} \bigg(\frac{k_1}{k_3},\frac{k_2}{k_3}\bigg)=&\mp\frac{(k_1k_2k_3)^2}{(2\pi)^4 P_\zeta^2}\text{Re}\left\{\left[\frac{8c_{2}c_{3}H^{5}}{(2\pi)^{13/2}\epsilon^{3}M^{6}_{P}k_{1}k_{2}(k_{1}+k_{2})^{4}}\right]\left[\frac{k_{3}}{4(k_{1}+k_{2})}\right]^{2i\mu}
g(\mu)\right\}\nonumber\\
&+(\text{even permutations in $k_{1},k_{2},k_{3}$})\label{eq:shape}
\end{align}
In the squeezed limit $k_{1}\sim k_{2},k_{1,2}\gg k_{3}$. The oscillatory term that is of interest to us comes from the first term of~(\ref{eq:shape}) as shown in fig.~\ref{fig:genericS}.
\section{Conclusion and outlook}
\label{conclusion}
In this paper, we have developed the in-in and SK formalism for pseudo Hermitian field theories and applied them to study the inflationary space-time. We find, as long as the pseudo Hermitian Hamiltonian has real spectrum, the correlators can be computed in the same way as Hermitian theories.


To illustrate the formalism, we use our method to study a model which consists of a massive symplectic fermions coupled to the primordial curvature perturbation. We calculate the three-point function of the primordial curvature perturbation using the SK formalism up to one-loop. We compare this to the previous computation where the one-loop correction comes from the massive scalar boson~\cite{Li:2019ves} and found they differ by a minus sign. While this minus sign is not observable, we can think of at least two ways to distinguish the scalar bosons and fermions. One way is to add new interactions between the matter fields and the primordial curvature perturbation. This is straightforward but not very interesting.

A more interesting possibility is to use the same models and compare the production rates of the symplectic fermions and scalar bosons at late times as $\tau\rightarrow0$ which can be derived from their respective Bogoliubov coefficients. In the bosonic case, it is known that the magnitude of the Bogoliubov coefficients yields the Bose-Einstein distribution~\cite{Li:2019ves}. Based on this result, it is reasonable to believe magnitude of the fermionic Bogoliubov coefficients would yield the Fermi-Dirac distribution.


Apart from correlators, there are other important observables in quantum field theory associated with the $S$-matrix (such cross-section and decay rates). While we have presented the formalism to compute correlators for pseudo Hermitian field theories, the formalism to extract observables from with the $S$-matrix has yet to be developed in Minkowski space-time. In our opinion, the challenge is to write down the formula for transition probability which is consistent with unitarity (or generalized unitarity) and the optical theorem. These are open problems for non-Hermitian theories. Works addressing these problems mostly concern non-Hermitian $PT$ symmetric Hamiltonians~\cite{Bender:1998ke,Bender:2004vn,Bender:2004sa,Bender:2007nj} and are not directly relevant to the pseduo Hermitian field theories considered here and in~\cite{LeClair_2007,Ahluwalia:2022zrm,Lee:2023aip,Sablevice:2023odu,Ahluwalia:2023slc}. Recent work on dS S-matrix and optical theorem~\cite{Donath:2024utn} indicates that these issues are also important in cosmology. We leave these important problems for future investigations.

\appendix 

\section{Propagators in the Schwinger Keldysh formalism}\label{sk_prop}

Here we compute $\langle\zeta_{\boldsymbol{k}_{1}}\zeta_{\boldsymbol{k}_{2}}\zeta_{\boldsymbol{k}_{3}}\rangle_{\pm}$. For most part, we focus on the fermionic three-point function, namely $\langle\zeta_{\boldsymbol{k}_{1}}\zeta_{\boldsymbol{k}_{2}}\zeta_{\boldsymbol{k}_{3}}\rangle_{+}$. The expression for its bosonic counterpart $\langle\zeta_{\boldsymbol{k}_{1}}\zeta_{\boldsymbol{k}_{2}}\zeta_{\boldsymbol{k}_{3}}\rangle_{-}$ can be obtained afterwards. 

To compute the three-point function, we need the propagators for the symplectic fermions and $\zeta$. In the configuration space, they are given by
\begin{equation}
    \left[\begin{matrix}
        D_{++}(\tau_{1},\x_{1};\tau_{2},\x_{2}) & D_{+-}(\tau_{1},\x_{1};\tau_{2},\x_{2}) \\
        D_{-+}(\tau_{1},\x_{1};\tau_{2},\x_{2}) & D_{--}(\tau_{1},\x_{1};\tau_{2},\x_{2})
    \end{matrix}\right]=
    \left[\begin{matrix}
        \langle T\sigma(\tau_{1},\x_{1})\dual{\sigma}(\tau_{2},\x_{2})\rangle & -\langle\dual{\sigma}(\tau_{2},\x_{2})\sigma(\tau_{1},\x_{1})\rangle \\
        \langle \sigma(\tau_{1},\x_{1})\dual{\sigma}(\tau_{2},\x_{2})\rangle &  -\langle\overline{T}\sigma(\tau_{1},\x_{1})\dual{\sigma}(\tau_{2},\x_{2})\rangle
    \end{matrix}\right],
\end{equation}
and
\begin{equation}
    \left[\begin{matrix}
        G_{++}(\k_{1},\tau_{1};\k_{2},\tau_{2}) & G_{+-}(\k_{1},\tau_{1};\k_{2},\tau_{2}) \\
        G_{-+}(\k_{1},\tau_{1};\k_{2},\tau_{2}) & G_{--}(\k_{1},\tau_{1};\k_{2},\tau_{2})
    \end{matrix}\right]=
    \left[\begin{matrix}
        \langle T\zeta(\k_{1},\tau_{1})\zeta(\k_{2},\tau_{2})\rangle & \langle\zeta(\k_{2},\tau_{2})\zeta(\k_{1},\tau_{1})\rangle \\
        \langle \zeta(\k_{1},\tau_{1})\zeta(\k_{2},\tau_{2})\rangle &  \langle\overline{T}\zeta(\k_{1},\tau_{1})\zeta(\k_{2},\tau_{2})\rangle
    \end{matrix}\right].
\end{equation}
The fields $\sigma$ and $\dual{\sigma}$ are fermionic so their time-order and anti time-order products are
\begin{align}
    D_{++}(\tau_{1},\x_{1};\tau_{2},\x_{2})&=\langle \sigma(\tau_{1},\x_{1})\dual{\sigma}(\tau_{2},\x_{2})\rangle\theta(\tau_{1}-\tau_{2})-\langle\dual{\sigma}(\tau_{2},\x_{2})\sigma(\tau_{1},\x_{1})\rangle\theta(\tau_{2}-\tau_{1}),\label{eq:dpp} \\
    D_{--}(\tau_{1},\x_{1};\tau_{2},\x_{2})&=\langle\dual{\sigma}(\tau_{2},\x_{2})\sigma(\tau_{1},\x_{1})\rangle\theta(\tau_{1}-\tau_{2})-\langle\sigma(\tau_{1},\x_{1})\dual{\sigma}(\tau_{2},\sigma_{2})\rangle\theta(\tau_{2}-\tau_{1}),\label{eq:dmm}
\end{align}
where $\theta$ is the step function. The propagators in the configuration and momentum space are related by
\begin{equation}
    D_{ab}(\tau_{1},\x_{1};\tau_{2},\x_{2})=\int\frac{d^{3}k}{(2\pi)^{3}}e^{i\boldsymbol{k\cdot(x_{1}-x_{2})}}D_{ab}(k;\tau_{1},\tau_{2}).
\end{equation}
Substituting the mode functions~(\ref{eq:phix_i}-\ref{eq:dsigma}) into~(\ref{eq:dpp}-\ref{eq:dmm}), we obtain
\begin{align}
 D_{++}(k;\tau_{1},\tau_{2})&=v_k\left(\tau_1\right) v_k^*\left(\tau_2\right) \theta\left(\tau_1-\tau_2\right)+v_k^*\left(\tau_1\right) v_k\left(\tau_2\right) \theta\left(\tau_2-\tau_1\right), \\
D_{--}(k;\tau_{1},\tau_{2})&=v_k^*\left(\tau_1\right) v_k\left(\tau_2\right) \theta\left(\tau_1-\tau_2\right)+v_k\left(\tau_1\right) v_k^*\left(\tau_2\right) \theta\left(\tau_2-\tau_1\right),  
\end{align}
and
\begin{align}
D_{+-}(k;\tau_{1},\tau_{2})&=v_k^*\left(\tau_1\right) v_k\left(\tau_2\right), \label{eq:dpm} \\
D_{-+}(k,\tau_{1},\tau_{2})&=v_k\left(\tau_1\right) v_k^*\left(\tau_2\right).
\end{align}
It is important to note that while the symplectic fermionic propagators are different to the bosonic scalar propagators in the configuration space, they become identical in the momentum space.

The primordial curvature perturbation is bosonic so
\begin{align}
    G_{++}(\tau_{1},\x_{1};\tau_{2},\x_{2})&=\langle \zeta(\tau_{1},\x_{1})\zeta(\tau_{2},\x_{2})\rangle\theta(\tau_{1}-\tau_{2})+\langle\zeta(\tau_{2},\x_{2})\zeta(\tau_{1},\x_{1})\rangle\theta(\tau_{2}-\tau_{1}), \\
    G_{--}(\tau_{1},\x_{1};\tau_{2},\x_{2})&=\langle\zeta(\tau_{2},\x_{2})\zeta(\tau_{1},\x_{1})\rangle\theta(\tau_{1}-\tau_{2})+\langle\zeta(\tau_{1},\x_{1})\zeta(\tau_{2},\x_{2})\rangle\theta(\tau_{2}-\tau_{1}).
\end{align}
Their propagators in momentum space are
\begin{align}
 G_{++}(k;\tau_{1};\tau_{2})&=u_k\left(\tau_1\right) u_k^*\left(\tau_2\right) \theta\left(\tau_1-\tau_2\right)+u_k^*\left(\tau_1\right) u_k\left(\tau_2\right) \theta\left(\tau_2-\tau_1\right),\label{eq:gpp} \\
 G_{+-}(k;\tau_{1},\tau_{2})&=u_k^*\left(\tau_1\right) u_k\left(\tau_2\right), \\
 G_{-+}(k;\tau_{1},\tau_{2})&=u_k\left(\tau_1\right) u_k^*\left(\tau_2\right), \\
 G_{--}(k;\tau_{1},\tau_{2})&=u_k^*\left(\tau_1\right) u_k\left(\tau_2\right) \theta\left(\tau_1-\tau_2\right)+u_k\left(\tau_1\right) u_k^*\left(\tau_2\right) \theta\left(\tau_2-\tau_1\right).\label{eq:gmm}
\end{align}

Having obtained the propagators, we can now compute the three-point function. Using the SK Feynman rules derived in~\cite{Chen:2017ryl} and 
\begin{equation}
\left\langle\zeta_{\mathbf{k}_1} \zeta_{\mathbf{k}_2} \zeta_{\mathbf{k}_3}\right\rangle_{\pm}\equiv
\left\langle\zeta_{\mathbf{k}_1} \zeta_{\mathbf{k}_2} \zeta_{\mathbf{k}_3}\right\rangle_{\pm}'\left[(2\pi)^{3}\delta^{3}(\k_{1}+\k_{2}+\k_{3})\right],
\end{equation}
we obtain
\begin{align}
&\left\langle\zeta_{\mathbf{k}_1} \zeta_{\mathbf{k}_2} \zeta_{\mathbf{k}_3}\right\rangle_{\pm}'\nonumber\\
=&2c_3 c_4 \sum_{a, b= \pm} a b \int_{-\infty}^0 \int_{-\infty}^0 \frac{d \tau_1}{\left(-H \tau_1\right)^3} \frac{d \tau_2}{\left(-H \tau_2\right)^2}\int\frac{d^{3}p}{(2 \pi)^3}\int\frac{d^{3}q}{(2 \pi)^3}\delta^{3}(\p+\q-\k_{3})\nonumber\\
&\times\left[\partial_{\tau_1} G_{a+}\left(k_3, \tau_1, 0\right)\partial_{\tau_2} G_{b+}\left(k_1, \tau_2, 0\right) \partial_{\tau_2} G_{b+}\left(k_2, \tau_2, 0\right)+(\text{even permutations in $k_1,k_2,k_3$})\right]\nonumber\\
&\times\left[(\mp1)D_{a b}\left(p, \tau_1, \tau_2\right) D_{b a}\left(q, \tau_2, \tau_1\right)\right]\nonumber\\
=&4c_3 c_4 \text{Re}\int_{-\infty}^0 \int_{-\infty}^0 \frac{d \tau_1}{\left(-H \tau_1\right)^3} \frac{d \tau_2}{\left(-H \tau_2\right)^2}\int\frac{d^{3}p}{(2 \pi)^3}\int\frac{d^{3}q}{(2 \pi)^3}\delta^{3}(\p+\q-\k_{3})\nonumber\\
&\times\Big\{\left[\partial_{\tau_{1}}G_{++}(k_{3},\tau_{1},0)+\partial_{\tau_{1}}G_{-+}(k_{3},\tau_{1},0)\right]
\left[\mp1 D(p,\tau_{1},\tau_{2})D(q,\tau_{2},\tau_{1})\right]\nonumber\\
&\times\left[\partial_{\tau_{2}}G_{++}(k_{1},\tau_{2},0)\partial_{\tau_{2}}G_{++}(k_{2},\tau_{2},0)\right]+(\text{even permutations in $k_1,k_2,k_3$})\Big\}.\label{eq:z3k}
\end{align}
In~(\ref{eq:z3k}), the $\mp1$ phase is due to the statistics of the fields. The propagators $G_{ab}$ can be further simplified. Using~(\ref{eq:gpp}-\ref{eq:gmm}), we find
\begin{align}
    G_{++}(k_{3},\tau_{1},0)&=u^{*}_{k_{3}}(\tau_{1})u_{k_{3}}(0),\label{eq:gp} \\
    G_{-+}(k_{3},\tau_{1},0)&=u_{k_{3}}(\tau_{1})u_{k_{3}}(0),\label{eq:gm}
\end{align}
where we have used the identity $u^{*}_{k}(0)=u_{k}(0)$. Substituting~(\ref{eq:gp}-\ref{eq:gm}) into~(\ref{eq:z3k}), we obtain
\begin{align}
&\left\langle\zeta_{\mathbf{k}_1} \zeta_{\mathbf{k}_2} \zeta_{\mathbf{k}_3}\right\rangle_{\pm}'\nonumber\\
=&4c_3 c_4 \text{Re}\int_{-\infty}^0 \int_{-\infty}^0 \frac{d \tau_1}{\left(-H \tau_1\right)^3} \frac{d \tau_2}{\left(-H \tau_2\right)^2}\int\frac{d^{3}p}{(2 \pi)^3}\int\frac{d^{3}q}{(2 \pi)^3}\delta^{3}(\p+\q-\k_{3})u_{k_{1}}(0)u_{k_{2}}(0)u_{k_{3}}(0)\nonumber\\
&\times\Big\{\left[\partial_{\tau_{1}}u^{*}_{k_{3}}(\tau_{1})+\partial_{\tau_{1}}u_{k_{3}}(\tau_{1})\right]
\left[\mp1 D(p,\tau_{1},\tau_{2})D(q,\tau_{2},\tau_{1})\right]
\left[\partial_{\tau_{2}}u^{*}_{k_{1}}(\tau_{2})\partial_{\tau_{2}}u^{*}_{k_{2}}(\tau_{2})\right]\Big\}\nonumber\\
&+(\text{even permutations in $k_1,k_2,k_3$}).
\end{align}
As we are interested in the long-distance correlation, we take the squeezed limit in which all the propagators become identical~\cite{Arkani-Hamed:2015bza,Lee_2016,Li:2019ves}
\begin{equation}
    D\equiv D_{++}=D_{--}=D_{+-}=D_{-+},
\end{equation}
where
\begin{equation}
    D(k,\tau_{1},\tau_{2})=\frac{a^{-3/2}(\tau_{1})a^{-3/2}(\tau_{2})}{4\pi H}
    \left[\Gamma^{2}(-i\mu)\left(\frac{k^{2}\tau_{1}\tau_{2}}{4}\right)^{i\mu}+\text{c.c.}\right]+\cdots.\label{eq:d_sq}
\end{equation}
Substituting~(\ref{eq:d_sq}) into~(\ref{eq:z3k}) we obtain
\begin{align}
\left\langle\zeta_{\mathbf{k}_1} \zeta_{\mathbf{k}_2} \zeta_{\mathbf{k}_3}\right\rangle'_{\pm}
=-\frac{c_{3}c_{4}}{4\pi^{2}H^{2}}\text{Re}\Big\{&\int^{0}_{-\infty}d\tau_{1}\int^{0}_{-\infty}d\tau_{2}\int\frac{d^{3}p}{(2\pi)^3}\int\frac{d^3q}{(2\pi)^3}\nonumber\\
\times&\delta^{3}(\p+\q-\k_{3})u(k_{1},0)u(k_{2},0)u(k_{3},0)\nonumber\\
\times&\left[\partial_{\tau_1}u^{*}(k_{3},\tau_{1})-\partial_{\tau_{1}}u(k_{3},\tau_{1})\right]
(\mp 1)\left[\Gamma^{2}(-i\mu)\left(\frac{pq\tau_{1}}{4}\right)^{2i\mu}\tau^{1+2i\mu}_{2}+\text{c.c.}\right]\nonumber\\
\times&\left[\partial_{\tau_2}u^{*}\left(k_1,\tau_2\right)\right]\left[\partial_{\tau_{2}}u^{*}\left(k_{2},\tau_{2}\right)\right]+(\text{even permutations in $k_{1},k_{2},k_{3}$})\Big\}.
\end{align}

\acknowledgments

We would like to thank Guoen Nian for the discussion in the early stage of this work. We are grateful to Yi Wang for useful discussions. SZ is supported by Natural Science Foundation of China under Grant No.12347101, No.2024CDJXY022 and No.02330052020052 at Chongqing University. LL is supported by NSFC grant
NO. 12165009, Hunan Natural Science Foundation NO. 2023JJ30487 and NO. 2022JJ40340.

\bibliography{Bibliography}

\providecommand{\href}[2]{#2}\begingroup\raggedright\begin{thebibliography}{10}

\bibitem{Ashida:2020dkc}
Y.~Ashida, Z.~Gong and M.~Ueda, \emph{{Non-Hermitian physics}},
  \href{http://dx.doi.org/10.1080/00018732.2021.1876991}{\emph{Adv. Phys.} {\bf
  69} (2021) 249--435}, [\href{https://arxiv.org/abs/2006.01837}{{\tt
  2006.01837}}].

\bibitem{Liu:2018kfw}
H.~Liu and P.~Glorioso, \emph{{Lectures on non-equilibrium effective field
  theories and fluctuating hydrodynamics}},
  \href{http://dx.doi.org/10.22323/1.305.0008}{\emph{PoS} {\bf TASI2017} (2018)
  008}, [\href{https://arxiv.org/abs/1805.09331}{{\tt 1805.09331}}].

\bibitem{Brauner:2022rvf}
T.~Brauner, S.~A. Hartnoll, P.~Kovtun, H.~Liu, M.~Mezei, A.~Nicolis et~al.,
  \emph{{Snowmass White Paper: Effective Field Theories for Condensed Matter
  Systems}},  in \emph{{Snowmass 2021}}, 3, 2022.
\newblock \href{https://arxiv.org/abs/2203.10110}{{\tt 2203.10110}}.

\bibitem{Hongo:2018ant}
M.~Hongo, S.~Kim, T.~Noumi and A.~Ota, \emph{{Effective field theory of
  time-translational symmetry breaking in nonequilibrium open system}},
  \href{http://dx.doi.org/10.1007/JHEP02(2019)131}{\emph{JHEP} {\bf 02} (2019)
  131}, [\href{https://arxiv.org/abs/1805.06240}{{\tt 1805.06240}}].

\bibitem{Salcedo:2024smn}
S.~A. Salcedo, T.~Colas and E.~Pajer, \emph{{The Open Effective Field Theory of
  Inflation}},  \href{https://arxiv.org/abs/2404.15416}{{\tt 2404.15416}}.

\bibitem{Mostafazadeh:2001jk}
A.~Mostafazadeh, \emph{{PseudoHermiticity versus PT symmetry. The necessary
  condition for the reality of the spectrum}},
  \href{http://dx.doi.org/10.1063/1.1418246}{\emph{J. Math. Phys.} {\bf 43}
  (2002) 205--214}, [\href{https://arxiv.org/abs/math-ph/0107001}{{\tt
  math-ph/0107001}}].

\bibitem{Mostafazadeh:2008pw}
A.~Mostafazadeh, \emph{{Pseudo-Hermitian Representation of Quantum Mechanics}},
  \href{http://dx.doi.org/10.1142/S0219887810004816}{\emph{Int. J. Geom. Meth.
  Mod. Phys.} {\bf 7} (2010) 1191--1306},
  [\href{https://arxiv.org/abs/0810.5643}{{\tt 0810.5643}}].

\bibitem{Kapit:2008dp}
E.~Kapit and A.~LeClair, \emph{{A Model of a 2d non-Fermi liquid with SO(5)
  symmetry, AF order, and a d-wave SC gap}},
  \href{http://dx.doi.org/10.1088/1751-8113/42/2/025402}{\emph{J. Phys. A} {\bf
  42} (2009) 025402}, [\href{https://arxiv.org/abs/0805.4182}{{\tt
  0805.4182}}].

\bibitem{LeClair_2007}
A.~LeClair and M.~Neubert, \emph{Semi-lorentz invariance, unitarity, and
  critical exponents of symplectic fermion models},
  \href{http://dx.doi.org/10.1088/1126-6708/2007/10/027}{\emph{Journal of High
  Energy Physics} {\bf 2007} (oct, 2007) 027}.

\bibitem{Fei:2015kta}
L.~Fei, S.~Giombi, I.~R. Klebanov and G.~Tarnopolsky, \emph{{Critical Sp(N )
  models in 6 \ensuremath{-} \ensuremath{\epsilon} dimensions and higher spin
  dS/CFT}}, \href{http://dx.doi.org/10.1007/JHEP09(2015)076}{\emph{JHEP} {\bf
  09} (2015) 076}, [\href{https://arxiv.org/abs/1502.07271}{{\tt 1502.07271}}].

\bibitem{Sato:2015tta}
Y.~Sato, \emph{{Comments on Entanglement Entropy in the dS/CFT
  Correspondence}},
  \href{http://dx.doi.org/10.1103/PhysRevD.91.086009}{\emph{Phys. Rev. D} {\bf
  91} (2015) 086009}, [\href{https://arxiv.org/abs/1501.04903}{{\tt
  1501.04903}}].

\bibitem{Anninos:2011ui}
D.~Anninos, T.~Hartman and A.~Strominger, \emph{{Higher Spin Realization of the
  dS/CFT Correspondence}},
  \href{http://dx.doi.org/10.1088/1361-6382/34/1/015009}{\emph{Class. Quant.
  Grav.} {\bf 34} (2017) 015009}, [\href{https://arxiv.org/abs/1108.5735}{{\tt
  1108.5735}}].

\bibitem{Ng:2012xp}
G.~S. Ng and A.~Strominger, \emph{{State/Operator Correspondence in Higher-Spin
  dS/CFT}},
  \href{http://dx.doi.org/10.1088/0264-9381/30/10/104002}{\emph{Class. Quant.
  Grav.} {\bf 30} (2013) 104002}, [\href{https://arxiv.org/abs/1204.1057}{{\tt
  1204.1057}}].

\bibitem{Chang:2013afa}
C.-M. Chang, A.~Pathak and A.~Strominger, \emph{{Non-Minimal Higher-Spin
  DS4/CFT3}},  \href{https://arxiv.org/abs/1309.7413}{{\tt 1309.7413}}.

\bibitem{Ryu:2022ffg}
S.~Ryu and J.~Yoon, \emph{{Unitarity of Symplectic Fermions in
  \ensuremath{\alpha} Vacua with Negative Central Charge}},
  \href{http://dx.doi.org/10.1103/PhysRevLett.130.241602}{\emph{Phys. Rev.
  Lett.} {\bf 130} (2023) 241602},
  [\href{https://arxiv.org/abs/2208.12169}{{\tt 2208.12169}}].

\bibitem{Alexandre:2019jdb}
J.~Alexandre, J.~Ellis, P.~Millington and D.~Seynaeve, \emph{{Spontaneously
  Breaking Non-Abelian Gauge Symmetry in Non-Hermitian Field Theories}},
  \href{http://dx.doi.org/10.1103/PhysRevD.101.035008}{\emph{Phys. Rev. D} {\bf
  101} (2020) 035008}, [\href{https://arxiv.org/abs/1910.03985}{{\tt
  1910.03985}}].

\bibitem{Alexandre:2020wki}
J.~Alexandre, J.~Ellis and P.~Millington, \emph{{$\mathcal{PT}$-symmetric
  non-Hermitian quantum field theories with supersymmetry}},
  \href{http://dx.doi.org/10.1103/PhysRevD.101.085015}{\emph{Phys. Rev. D} {\bf
  101} (2020) 085015}, [\href{https://arxiv.org/abs/2001.11996}{{\tt
  2001.11996}}].

\bibitem{Alexandre:2020gah}
J.~Alexandre, J.~Ellis and P.~Millington, \emph{{Discrete spacetime symmetries
  and particle mixing in non-Hermitian scalar quantum field theories}},
  \href{http://dx.doi.org/10.1103/PhysRevD.102.125030}{\emph{Phys. Rev. D} {\bf
  102} (2020) 125030}, [\href{https://arxiv.org/abs/2006.06656}{{\tt
  2006.06656}}].

\bibitem{Alexandre:2022uns}
J.~Alexandre, J.~Ellis and P.~Millington, \emph{{Discrete spacetime symmetries,
  second quantization, and inner products in a non-Hermitian Dirac fermionic
  field theory}},
  \href{http://dx.doi.org/10.1103/PhysRevD.106.065003}{\emph{Phys. Rev. D} {\bf
  106} (2022) 065003}, [\href{https://arxiv.org/abs/2201.11061}{{\tt
  2201.11061}}].

\bibitem{Mason:2023sgr}
R.~Mason, P.~Millington and E.~Sablevice, \emph{{Flavour oscillations in
  pseudo-Hermitian quantum theories}},
  \href{http://dx.doi.org/10.22323/1.449.0498}{\emph{PoS} {\bf EPS-HEP2023}
  (2024) 498}, [\href{https://arxiv.org/abs/2311.04839}{{\tt 2311.04839}}].

\bibitem{Alexandre:2023afi}
J.~Alexandre, M.~Dale, J.~Ellis, R.~Mason and P.~Millington, \emph{{Oscillation
  probabilities for a $\mathcal{PT}$-symmetric non-Hermitian two-state
  system}},  \href{https://arxiv.org/abs/2302.11666}{{\tt 2302.11666}}.

\bibitem{Ahluwalia:2022zrm}
D.~V. Ahluwalia and C.-Y. Lee, \emph{{Spin-half bosons with mass dimension
  three-half: Evading the spin-statistics theorem}},
  \href{http://dx.doi.org/10.1209/0295-5075/ac97bd}{\emph{EPL} {\bf 140} (2022)
  24001}, [\href{https://arxiv.org/abs/2212.09457}{{\tt 2212.09457}}].

\bibitem{Lee:2023aip}
C.-Y. Lee, \emph{{Generalized unitary evolution for symplectic scalar
  fermions}}, \href{http://dx.doi.org/10.1007/JHEP05(2024)181}{\emph{JHEP} {\bf
  05} (2024) 181}, [\href{https://arxiv.org/abs/2305.17712}{{\tt 2305.17712}}].

\bibitem{Sablevice:2023odu}
E.~Sablevice and P.~Millington, \emph{{Poincar\'e symmetries and
  representations in pseudo-Hermitian quantum field theory}},
  \href{http://dx.doi.org/10.1103/PhysRevD.109.065012}{\emph{Phys. Rev. D} {\bf
  109} (2024) 065012}, [\href{https://arxiv.org/abs/2307.16805}{{\tt
  2307.16805}}].

\bibitem{Ahluwalia:2023slc}
D.~V. Ahluwalia, G.~B. de~Gracia, J.~M.~H. da~Silva, C.-Y. Lee and B.~M.
  Pimentel, \emph{{Irreducible representations of the inhomogeneous Lorentz
  group with two-fold Wigner degeneracy}},
  \href{https://arxiv.org/abs/2312.17038}{{\tt 2312.17038}}.

\bibitem{Schwinger:1960qe}
J.~S. Schwinger, \emph{{Brownian motion of a quantum oscillator}},
  \href{http://dx.doi.org/10.1063/1.1703727}{\emph{J. Math. Phys.} {\bf 2}
  (1961) 407--432}.

\bibitem{PhysRevD.33.444}
R.~D. Jordan, \emph{Effective field equations for expectation values},
  \href{http://dx.doi.org/10.1103/PhysRevD.33.444}{\emph{Phys. Rev. D} {\bf 33}
  (Jan, 1986) 444--454}.

\bibitem{Weinberg:2005vy}
S.~Weinberg, \emph{{Quantum contributions to cosmological correlations}},
  \href{http://dx.doi.org/10.1103/PhysRevD.72.043514}{\emph{Phys. Rev. D} {\bf
  72} (2005) 043514}, [\href{https://arxiv.org/abs/hep-th/0506236}{{\tt
  hep-th/0506236}}].

\bibitem{Chen:2016nrs}
X.~Chen, Y.~Wang and Z.-Z. Xianyu, \emph{{Loop Corrections to Standard Model
  Fields in Inflation}},
  \href{http://dx.doi.org/10.1007/JHEP08(2016)051}{\emph{JHEP} {\bf 08} (2016)
  051}, [\href{https://arxiv.org/abs/1604.07841}{{\tt 1604.07841}}].

\bibitem{Chen:2017ryl}
X.~Chen, Y.~Wang and Z.-Z. Xianyu, \emph{{Schwinger-Keldysh Diagrammatics for
  Primordial Perturbations}},
  \href{http://dx.doi.org/10.1088/1475-7516/2017/12/006}{\emph{JCAP} {\bf 12}
  (2017) 006}, [\href{https://arxiv.org/abs/1703.10166}{{\tt 1703.10166}}].

\bibitem{Donath:2024utn}
Y.~Donath and E.~Pajer, \emph{{The in-out formalism for in-in correlators}},
  \href{http://dx.doi.org/10.1007/JHEP07(2024)064}{\emph{JHEP} {\bf 07} (2024)
  064}, [\href{https://arxiv.org/abs/2402.05999}{{\tt 2402.05999}}].

\bibitem{Wang:2013zva}
Y.~Wang, \emph{{Inflation, Cosmic Perturbations and Non-Gaussianities}},
  \href{http://dx.doi.org/10.1088/0253-6102/62/1/19}{\emph{Commun. Theor.
  Phys.} {\bf 62} (2014) 109--166},
  [\href{https://arxiv.org/abs/1303.1523}{{\tt 1303.1523}}].

\bibitem{Li:2019ves}
L.~Li, T.~Nakama, C.~M. Sou, Y.~Wang and S.~Zhou, \emph{{Gravitational
  Production of Superheavy Dark Matter and Associated Cosmological
  Signatures}}, \href{http://dx.doi.org/10.1007/JHEP07(2019)067}{\emph{JHEP}
  {\bf 07} (2019) 067}, [\href{https://arxiv.org/abs/1903.08842}{{\tt
  1903.08842}}].

\bibitem{Lee:2024}
T.-L. Feng, C.-Y. Lee, L.-H. Liu and S.~Zhou, \emph{{In preparation}}, .

\bibitem{Chen:2010xka}
X.~Chen, \emph{{Primordial Non-Gaussianities from Inflation Models}},
  \href{http://dx.doi.org/10.1155/2010/638979}{\emph{Adv. Astron.} {\bf 2010}
  (2010) 638979}, [\href{https://arxiv.org/abs/1002.1416}{{\tt 1002.1416}}].

\bibitem{Gelfand:1964}
I.~M. Gel'fand and G.~E. Shilov, \emph{{Generalized functions. Volume 1}}.
\newblock Academic Press Inc., 1964,
  \href{http://dx.doi.org/10.1017/9781316145593}{10.1017/9781316145593}.

\bibitem{Bender:1998ke}
C.~M. Bender and S.~Boettcher, \emph{{Real spectra in nonHermitian Hamiltonians
  having PT symmetry}},
  \href{http://dx.doi.org/10.1103/PhysRevLett.80.5243}{\emph{Phys. Rev. Lett.}
  {\bf 80} (1998) 5243--5246},
  [\href{https://arxiv.org/abs/physics/9712001}{{\tt physics/9712001}}].

\bibitem{Bender:2004vn}
C.~M. Bender, D.~C. Brody and H.~F. Jones, \emph{{Scalar quantum field theory
  with cubic interaction}},
  \href{http://dx.doi.org/10.1103/PhysRevLett.93.251601}{\emph{Phys. Rev.
  Lett.} {\bf 93} (2004) 251601},
  [\href{https://arxiv.org/abs/hep-th/0402011}{{\tt hep-th/0402011}}].

\bibitem{Bender:2004sa}
C.~M. Bender, D.~C. Brody and H.~F. Jones, \emph{{Extension of PT symmetric
  quantum mechanics to quantum field theory with cubic interaction}},
  \href{http://dx.doi.org/10.1103/PhysRevD.70.025001}{\emph{Phys. Rev. D} {\bf
  70} (2004) 025001}, [\href{https://arxiv.org/abs/hep-th/0402183}{{\tt
  hep-th/0402183}}].

\bibitem{Bender:2007nj}
C.~M. Bender, \emph{{Making sense of non-Hermitian Hamiltonians}},
  \href{http://dx.doi.org/10.1088/0034-4885/70/6/R03}{\emph{Rept. Prog. Phys.}
  {\bf 70} (2007) 947}, [\href{https://arxiv.org/abs/hep-th/0703096}{{\tt
  hep-th/0703096}}].

\bibitem{Arkani-Hamed:2015bza}
N.~Arkani-Hamed and J.~Maldacena, \emph{{Cosmological Collider Physics}},
  \href{https://arxiv.org/abs/1503.08043}{{\tt 1503.08043}}.

\bibitem{Lee_2016}
H.~Lee, D.~Baumann and G.~L. Pimentel, \emph{Non-gaussianity as a particle
  detector}, \href{http://dx.doi.org/10.1007/jhep12(2016)040}{\emph{Journal of
  High Energy Physics} {\bf 2016} (Dec., 2016) }.

\end{thebibliography}\endgroup
\bibliographystyle{JHEP}
\end{document}